\documentclass[reprint,
superscriptaddress,
amsmath,amssymb,
aps,prl,
twocolumn,
showpacs,
]{revtex4-1}
\usepackage{tikz}
\usepackage{graphicx}
\usepackage{dcolumn}
\usepackage{bm}
\usepackage{braket}
\usepackage{hyperref}
\usepackage{amsmath}
\usepackage{fancyhdr}
\usepackage{float}
\usepackage{todonotes}
\usepackage{xcolor}
\usepackage{xspace}
\usepackage{eucal}
\usepackage{ulem}
\usepackage{todonotes}
\usepackage{multirow}
\usepackage[utf8]{inputenc}
\usepackage{pgfplots}
\pgfplotsset{compat=1.14}
\usepackage{upgreek}
\usepackage{siunitx}
\usepackage{gensymb}
\usepackage{textcomp}
\usepackage{comment}
\usepackage{siunitx}
\usepackage{todonotes}
\usepackage{cleveref}
\usepackage{xfrac}
\usepackage{braket}
\usepackage{xspace}
\usepackage{siunitx}
\DeclareSIUnit\g{g}
\DeclareSIUnit\gal{Gal}
\DeclareSIUnit\torr{Torr}
\DeclareSIUnit\inch{inch}
\DeclareSIUnit\joule{J}
\newcommand{\Rb}{\textsuperscript{87}Rb\xspace}
\DeclareSIUnit{\inch}{''}
\DeclareSIUnit{\atoms}{atoms}
\DeclareSIUnit{\torr}{Torr}
\begin{document}
\title{Quantum hybrid optomechanical inertial sensing}
\author{Logan Richardson}
\author{Adam Hines}
\author{Andrew Schaffer}
\author{Brian P. Anderson}
\author{Felipe~Guzman}\email[Electronic mail: ]{felipe@optics.arizona.edu}
\affiliation{James C. Wyant College of Optical Sciences, University of Arizona, 1630 E. University Blvd., Tucson, AZ 85721, USA}%
\date{\today}
%
\begin{abstract}
We discuss the design of quantum hybrid inertial sensor that combines an optomechanical inertial sensor  with the retro-reflector of a cold atom interferometer. This sensor fusion approach provides absolute and high accuracy measurements with cold atom interferometers, while utilizing the optomechanical inertial sensor at frequencies above the repetition rate of the atom interferometer. This improves the overall measurement bandwidth as well as the robustness and field deployment capabilities of these systems. We evaluate which parameters yield an optimal acceleration sensitivity, from which we anticipate a noise floor at \textnormal{nano-}\emph{g} levels from DC to \SI{1}{\kilo\hertz}.
\end{abstract}
\maketitle
\section{Introduction}
When compared to drift-prone relative measurement devices, the accuracy and long-term stability offered by atom interferometers make them optimal devices for measuring accelerations in a wide field of applications ranging from gravimetry~\cite{freier2016mobile,bidel2018absolute,zhou2011development,farah2014underground,wu2019gravity} and civil engineering~\cite{hinton2017portable} to inertial navigation~\cite{garrido2019compact,cheiney2018navigation,doi:10.1116/1.5120348}. However, for many applications, the measurement bandwidth of atom interferometers, typically $0.25-\SI{4}{\hertz}$~\cite{AINATURE,chroninreview}, presents a real limitation. During the finite time required to measure acceleration with atom interferometers, retro-reflector dynamics with frequencies higher than the measurement cycle frequency ($f_c$), couple into the interferometer and manifest as inertial noise in the measured matter-wave phase~\cite{cheinet2008measurement}.  The bandwidth of atom interferometers is limited further by measurement dead-times, such as atom trap loading, state selection, and detection.

To address the effect of this external motion, efforts have been made to attenuate the high-frequency accelerations that couple into the system by using vibration isolation systems~\cite{gain,PhysRevA.86.043630}, or through post-correction with external classical sensors that track the retro-reflection mirror movement during measurement~\cite{Merlet_2009}. Vibration isolation systems, although effective in laboratory conditions are large, bulky and limited in dynamic range; which prohibits operation in extremely high-noise and in micro-$g$ environments . This frustrates efforts to reduce the size, weight, and power (SWAP) of atom interferometers for field use. Additionally, these approaches do not increase the measurement bandwidth of the sensor but instead focus on reducing the inertial noise. 

{Hybrid sensing offers a solution by combining the signal from a high bandwidth relative sensor with the absolute measurement of the atom interferometer, effectively creating a sensor with both high observation bandwidth and long-term stability~\cite{airbornegravimetry,cheiney2018navigation}. However, hybrid sensing requires accurate knowledge the mechanical transfer function between the relative sensor and atom interferometer signal, otherwise the hybrid sensor will suffer from reduced performance~\cite{bidel2020absolute}. This can be challenging if the parameters of this transfer function change dynamically.  Prototype electrostatic accelerometers~\cite{touboul1999electrostatic} which directly combine the test mass with the retro-reflector of the atom interferometer have been demonstrated~\cite{christophe2018status}, {however, such sensors are susceptible to external magnetic fields which are required to operate the atom interferometer, thus requiring additional shielding.} Previously, optomechanical sensors have been utilized to show atom interferometer post-correction in a proof-of-principle experiment~\cite{richardson2019opto}, however the sensors utilized here were not designed for atom interferometry and therefore suffered from less than ideal mechanical coupling to the atom interferometer retro-reflector. In this work we outline the design of a novel hybrid optomechanical inertial sensor, where the optomechanical element is intentionally designed for optimal integration into the atom interferometer.  This optimized sensor leads to a sensor fusion approach that makes the optomechanical element itself the retro-reflection mirror of the atom interferometer.} The optomechanical retro-reflector (OMRR) is a broadband inertial sensor made of fused silica, a low-loss non-magnetic material, and incorporates a compact and highly sensitive optical displacement sensor to read-out the position of its inertial sensing test mass. This test mass is also the mirror that reflects the laser beam utilized for atom interferometry. Having the OMRR test mass as the common inertial reference with the atom interferometer eliminates the effect from a mismatch in the mechanical overlap. Such a system will reduce the need for a vibration isolation system, and provides a path for more portable field-capable systems.

Optomechanical sensors have shown high sensitivity to changes in acceleration over a broad bandwidth up to \SI{10}{\kilo\hertz}~\cite{guzman2014high}. By careful selection of the materials, optomechanical sensors also offer advantages against stray electromagnetic fields, while being vacuum-compatible and allowing a high design control of their mechanical properties and performance. Inertial noise does not couple into the atom interferometer uniformly across all frequencies. Therefore, we can design the performance and frequency response of our OMRR for an optimal quantum hybrid inertial sensor. 

This work outlines the roadmap of our efforts to build such a quantum inertial sensor, as well as our design considerations for the required OMRR developments.

\section{Background}

\subsection{Atom interferometer}

Matter-wave interference of atoms enables repeatable precise measurement of inertial effects. Atom interferometry has been demonstrated with warm vapor, Bose-Einstein condensates, and in the case of this work, thermal cold atom clouds~\cite{AINATURE,chroninreview}.  We trap and cool an ensemble of atoms in a magneto-optical trap (MOT), and once this is loaded, the ensemble is released from the trap and falls along the direction of gravitational acceleration. By using timed pulses of light during free fall we can manipulate the matter-wave state of the atoms in such a manner that we can detect the acceleration they experienced through their corresponding phase shift. We chose to perform interferometry with \Rb for compatibility with other atom interferometer systems~\cite{richardson2019opto}, however, this measurement principle and OMRR technology can be applied to atom interferometry with other species.

During free-fall we use counter-propagating two-photon Raman pulses to drive transitions between the two hyperfine states $\ket{{F=1},m_\textnormal{F}=0}$ and $\ket{{F=2},m_\textnormal{F}=0}$ of the ${5} {}^{2}S_{1/2}$ energy level of \Rb, which we simplify to $\ket{1}$ and $\ket{2}$, respectively. We use these pulses to generate the matter-wave analogue of an optical beam-splitter ($\sfrac{\pi}{2}$ pulse) and mirror ($\pi$ pulse)~\cite{kasevich1992measurement}. By arranging these three pulses in a Mach-Zehnder-like configuration, $\sfrac{\pi}{2}-{\pi}-\sfrac{\pi}{2}$, with a separation time between pulses $T$, we create a matter-wave interferometer sensitive to inertial effects.

To meet the Raman condition required for inertially sensitive interferometry~\cite{kasevich1992measurement}, a two-beam counter-propagating configuration is realized by employing a retro-reflection mirror. For an interferometer initially prepared in the $\ket{1}$ state, the output population of the Mach-Zehnder-like intereferometer in the $\ket{2}$ state is given by:

\begin{equation}
    \label{eq:populationdetection}
    P_{\ket{2}} = \frac{C_0}{2}\Big[1+\cos\big(\Delta\Phi\big)\Big]+ B
\end{equation}
where $C_0$ is the contrast of the atom interferometer, $B$ is the offset and $\Delta \Phi$ is the total phase difference accumulated between the uncommon paths after the first ($\sfrac{\pi}{2}$) pulse. The population $P_{\ket{2}}$ is dependent on the phase difference, which for an atom interferometer aligned to measure gravitational acceleration is given by~\cite{kasevich1992measurement}:
\begin{equation}
\Delta \Phi = \vec{k}_\textnormal{eff} \cdot \vec{g} T^2   + \Phi_\textnormal{IN}  + \Phi_\textnormal{Other}
\label{eq:refname1}
\end{equation}


where $\vec{g}$ is the gravitational acceleration, $\vec{k}_\textnormal{eff}$ is the effective wave vector difference between the two counter-propagating Raman beams and  $\Phi_\textnormal{IN}$ is the phase shift given by the inertial noise experienced during interferometry, which we discuss further in Section~\ref{sec:vibrations}. {The term $\Phi_\textnormal{Other}$ consists of additional noise contributions such as laser phase noise~\cite{cheinet2008measurement}, detection noise, and laser intensity noise~\cite{le2008limits}.  These contributions are highly system specific, can be mitigated with careful interferometer design, and are typically not  the limiting factors for systems operating in high inertial noise environments.}

The fundamental limitation to the acceleration sensitivity of the atom interferometer $\sigma_a$ comes down to the ability to read the output population; ultimately determined by the quantum projection noise (QPN)~\cite{qpn}:
\begin{equation}
    \large{\sigma}_{a}^\textnormal{QPN} = \frac{1}{k_\textnormal{eff} T^2} \frac{1}{C_0 \sqrt{N}}
    \label{eq:qpn}
\end{equation}
where $N$ is the atom number. By designing the OMRR performance to reach the quantum projection noise limitation, we ensure that the quantum hybrid sensor sensitivity is fundamentally limited by the atom interferometer properties.  A lower $\sigma_a^\textnormal{QPN}$ will place more stringent requirements on the OMRR and therefore, we assume a full contrast level of one for this work. Similarly, we expect that the total atom number will decrease by an order of magnitude during sub-Doppler cooling~\cite{rosi2018lambda}, relaxing the OMRR requirements to reach the quantum projection noise limitation. However to maintain more rigorous constraints on the OMRR, we will assume that the atom number remains constant after loading the atoms into the magneto-optical trap. {Research is underway to improve the atom interferometer detection beyond the quantum projection noise, but these methods are beyond the scope of this work~\cite{beyondqpn}}.

\subsection{Vibrations and the atom interferometer}
\label{sec:vibrations}

Position fluctuations of the retro-reflection mirror $\delta {z}$ during interferometry will result in a wave-front phase shift, or phase jump, equivalent to $\delta \phi_\textnormal{IN} = \vec{{k}}_\textnormal{eff} \cdot \delta \vec{{z}}$. From the sensitivity formalism~\cite{cheinet2008measurement}, we obtain the sensitivity function $g(t)$ which weights how a phase jump, $\delta \phi$ at a given time $t$ during the atom interferometer cycle affects the total output population of the atom interferometer,
\begin{equation}
    g(t) = 2 \lim_{\delta\phi \to 0} \frac{\delta P_{\ket{2}} (\delta \phi,t)}{\delta \phi}.
    \label{eq:sensitivityfunction}
\end{equation}

By weighting phase shifts from the mirror displacement over the interferometer measurement cycle, we can obtain the total phase shift at the output of the atom interferometer,

\begin{equation}
    \Delta \Phi_\textnormal{IN} = \int^{\infty}_{-\infty} g(t) \, d\phi_\textnormal{\,IN}(t) =  \int^{\infty}_{-\infty} g(t) \, \frac{d\phi_\textnormal{\,IN}(t)}{dt} dt.
\end{equation}

If the change in mirror displacement is not measured during an atom interferometer cycle, it will manifest as an unknown phase shift, which we characterize as inertial noise. 

In frequency space, the finite measurement time of the atom interferometer will sample the inertial noise at intervals of $\omega_n = 2 \pi n f_c$. Given a background acceleration noise power spectral density, $S_a(2 \pi n f_c)$, over an averaging time, $\tau$, we can calculate the atom interferometer acceleration sensitivity limit, $\sigma_a$, for a set of $n$ measurements through~\cite{cheinet2008measurement}:

\begin{equation}
    \label{eq:sigmasq}
    \sigma_{a}^2 (\tau) = \frac{1}{\tau T^2} \sum_{n=0}^{\infty} \frac{\lvert H_{\phi}(2\pi n f_c)\rvert^2}{(2\pi n f_c)^4}  S_a(2\pi n f_c)
\end{equation}
where $H_\phi(\omega)$ is the atom interferometer transfer function, which describes its response to phase noise at a frequency $\omega$, and can be derived from the Fourier transform of the sensitivity function $g(t)$~\cite{cheinet2008measurement}.

In a quantum hybrid inertial sensor, we track the retro-reflector accelerations, and can therefore replace the external noise power spectral density $S_a(2 \pi n f_c)$ in Equation~\ref{eq:sigmasq} with the self-noise of the external sensor. The self-noise of an optomechanical sensor is determined primarily by the mechanical losses inherent to the resonator materials, geometry, operating environment, defined by pressure and temperature, and lastly by the displacement sensitivity limits to its test mass dynamics. Using the atom interferometer transfer function, we can design an optomechanical sensor with parameters that allow us to reach our intended hybrid sensor sensitivity.

\subsection{Optomechanical Sensors}

Our optomechanical sensors consist of mechanical oscillators that are monolithically fabricated from low loss materials. Low mechanical losses enable the oscillator's inertial sensing test mass to achieve very high sensitivities to external accelerations. We can observe these accelerations by measuring the test mass displacement through high-precision laser interferometry~\cite{guzman2014high}. {To this end, various techniques can be implemented such as optical resonant cavity read-out, or homodyne and heterodyne laser interferometry, which are discussed further in Section~3.\ref{sec:omrrdesign}}. The optical read-out of the test mass dynamics is not affected by external electromagnetic disturbances, which are a major concern when operating near the switching magnetic field coils required for atom interferometry. 

When determining the required design parameters for the targeted performance of an OMRR, we need to consider the bandwidth that is governed by the resonance, $\omega_0$, and the sensitivity of the sensor over that bandwidth. The displacement fluctuations of the inertial sensing test mass, $z(\omega)$, is ultimately limited by the mechanical losses in the resonator, which are given by:
\begin{equation}
    z(\omega) = \sqrt{ \frac{4k_{B}T_\textnormal{TM}k\phi(\omega)}{\omega((k-m\omega^{2})^{2}+k^{2}\phi(\omega)^{2})}} \label{dispPSD}
\end{equation}
where $T_\textnormal{TM}$ is the test mass temperature, $m$ is the effective mass of the test mass, $k_{B}$ is the Boltzmann constant, $k$ is the spring constant and $\phi(\omega)$ is the frequency-dependent loss-coefficient of the oscillator~\cite{Saulson}. The low-frequency behavior of the optomechanical sensor will be less critical to the hybrid sensor, as it will be referenced to the atom interferometer for frequencies below $f_c$. Therefore, the acceleration noise of the inertial test mass is dictated, in first order, by the acceleration thermal noise floor, $\delta a_{th}$, given by:
\begin{equation}
    \delta a_{th} = \sqrt{\frac{4k_{B}T_\textnormal{TM}\omega_{0}}{mQ}}
    \label{eq:thermalnoisefloor}
\end{equation}

where $\omega_0$ is the system resonance, and $Q$ is the mechanical quality factor.

As shown in Equation~\ref{eq:thermalnoisefloor}, we can lower the thermal noise by achieving high $mQ$-products in the mechanical oscillator as demonstrated in~\cite{OptomehanicalInertial,guzman2014high}. From these models, we can determine the required sensor resonance $\omega_0$, in order to measure inertial noise over a certain bandwidth, and the $mQ$-product to achieve the necessary sensitivity over this bandwidth for optimal hybridization with the atom interferometer.

Furthermore, test mass displacement amplitudes in our optomechanical sensors are typically very low under normal field accelerations, since their typical resonance frequencies resulting from the trade-off analysis outlined above, range from a few hundred Hz to a few kHz, leading to a high effective resonator stiffness. Hence, test mass displacement oscillations can be very well approximated to those of a simple harmonic oscillator, which translate to acceleration by the following transfer function~\cite{Saulson}:
\begin{equation}
    \label{eq:x2a}
    \frac{ z(\omega)}{ a(\omega) } = -\frac{1}{\omega_0^2-\omega^2+i\frac{\omega_0}{Q}\omega}
\end{equation}
where $z(\omega)$ is the displacement of the test mass at the Fourier angular frequency $\omega$ and $a(\omega)$ is the acceleration. 

\section{Design}

\subsection{Atom Interferometer Chamber Design}
Atom interferometer parameters such as the pulse separation time and cycle time ($T_c = \sfrac{1}{f_c}$) will help determine the optimal OMRR resonance $\omega_0$ and displacement sensitivity required. For this reason, we will briefly discuss the anticipated atom interferometer features pertinent to the design of the OMRR. 

We have designed a compact atom interferometer for quantum hybrid inertial sensing experiments. The design of this atom interferometer will allow for the rapid integration and testing of different optomechanical sensor designs. Light for cooling, trapping, and interferometry will be connected to MOT and interferometry telescopes via fiber optic cables.

\begin{figure}[htp]
\centering
\fbox{\includegraphics[width=.90\linewidth]{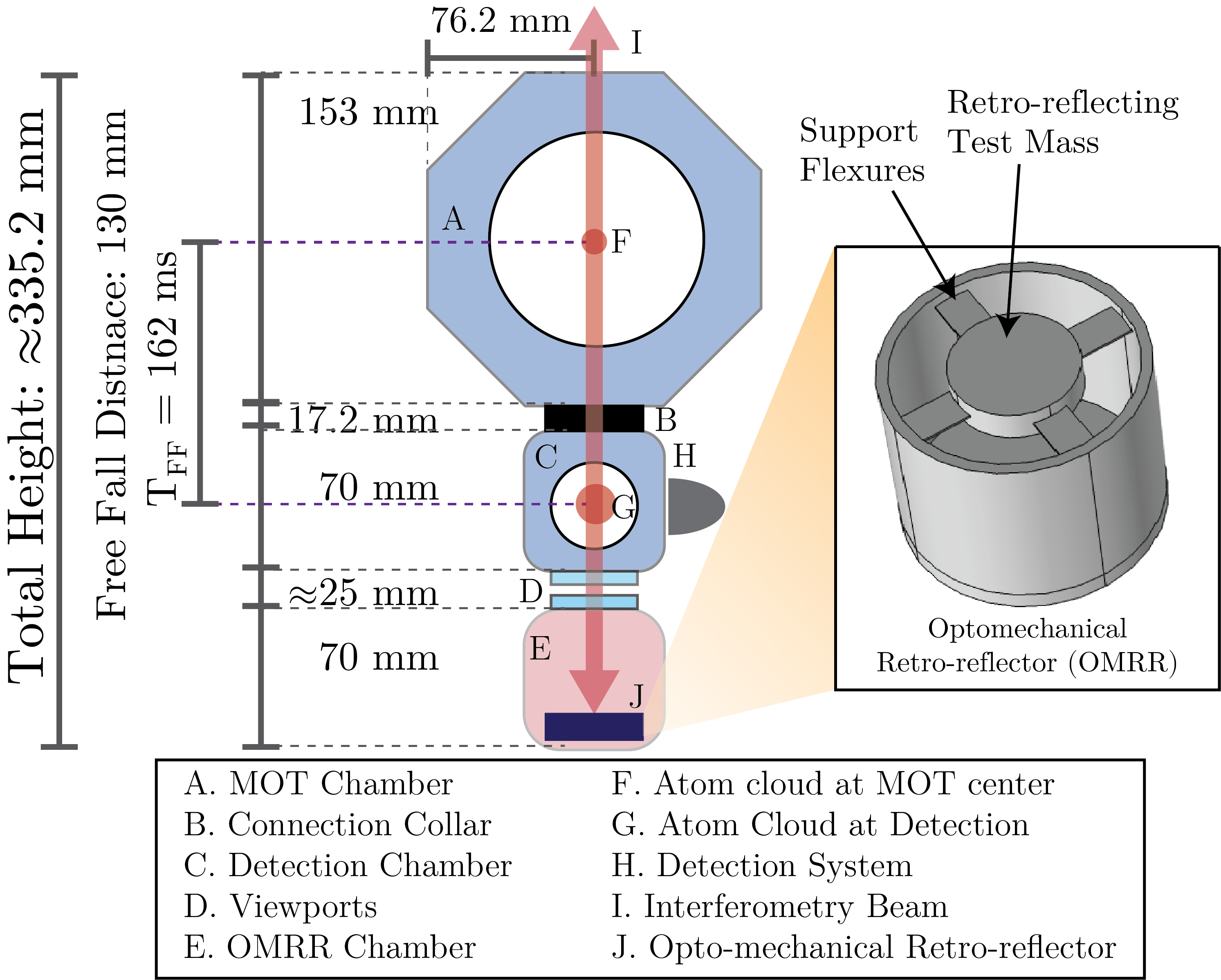}}
\caption{A diagram of the proposed testbed interferometer. The atom interferometer will consist of two separate vacuum systems at two different pressures; one for atom interferometry maintained at $\rho_a \leq \SI{7.5E-8}{\torr}$ and one for the OMRR maintained at $\rho_b \lesssim \SI{1E-4}{\torr}$. The total free fall distance from the center of the MOT to the detection region is roughly \SI{130}{\milli\meter} corresponding to a total free fall time of $\textnormal{T}_\textnormal{FF}=\SI{162}{\milli\second}$. The lower vacuum chamber will house the OMRR. A cutaway of the proposed OMRR can be seen depicting the retro-reflecting test mass and support flexures. {Atoms are cooled and trapped at the center of the MOT chamber (F) and once loading is complete, they are released into free fall. The interferometry beam (I) enters the system through the top of the MOT chamber (A) and reflected off of the OMRR (J) along the axis of gravitational acceleration, generating a counter-propagating configuration which satisfies conditions required for inertially sensitive interferometry. During free fall, three interferometry pulses separated by a pulse separation time $T$ are used to generate a Mach-Zehnder-like interferometer. Once the atoms reach the detection area (G), the phase-dependent (Equation~\ref{eq:populationdetection}) relative output population of the two state system is measured, from which the acceleration of the atoms can be determined.  Although  acceleration is discretely measured by the atom interferometer only once per cycle $T_\textnormal{c}$, acceleration as measured by the OMRR is continuous. Signal from the OMRR can be used to correct for vibrations occurring during atom interferometry, and  the absolute measurement of the atom interferometer can be used to debias the OMRR signal at frequencies below the atom interferometer cycle rate.}}
\label{fig:AIDiagram}
\end{figure}

The atom interferometer sensor head will consist of two separate vacuum chambers, one for atom interferometry and the other for the OMRR. Maintaining a separate vacuum system for the OMRR will allow us to easily access and rapidly test different OMRR designs without the need of breaking vacuum and baking out the atom interferometry chamber each time.  Both systems will be constructed of non-magnetic Unitary 316L stainless steel and are rigidly attached to one another. Both chambers will share a common optical axis that will allow retro-reflection of the atom interferometer beam through the use of viewports, as seen in Figure~\ref{fig:AIDiagram}. 

Atom cloud cooling, trapping, and interferometry will take place in the upper vacuum system in an octagonal chamber with eight DN35CF (\SI{2.75}{\inch} OD) Conflat ports  and two DN100CF (\SI{6}{\inch} OD) Conflat ports. After the MOT is loaded, the atom ensemble will be released from the trap and eventually will fall into a cube chamber connected below the octagonal chamber. This cube chamber has six DN35CF ports that we will use for detection, and will connect to an ion--getter pump. {The target pressure of the atom interferometry chamber must be sufficiently low such that collisions during interferometry will be minimal, experimentally this has been determined to be below $\rho_a \leq \SI{7.5E-8}{\torr}$}~\cite{RBMOTloading}. 

To reduce the effect of gas damping, we will place the OMRR into a separate six DN35CF port cube that will maintain a pressure $\rho_b \lesssim \SI{1E-4}{\torr}$, {a value which reduces the gas damping losses in this element to a negligible level as experimentally determined from similarly built sensors \cite{OptomehanicalInertial}}. Atom interferometry light from the upper chamber will pass through the bottom viewport into the cube chamber of the lower vacuum system. We will support the OMRR by an externally controllable positionable vacuum compatible mirror mount allowing for alignment of the interferometry axis parallel to gravity. 
 
For cooling and trapping, we will utilize six \SI{9}{\milli\meter} diameter optical beams, four of which will pass through the DN35CF ports, and two from the DN100CF viewports. We will load \Rb directly into the MOT from a natural abundance dispenser. This arrangement will allow for trapping in the center of the octagonal chamber, and gives a total free-fall distance of approximately \SI{130}{\milli\meter}, yielding a total drop time of $T_\textnormal{FF}=\SI{162}{\milli\second}$. Allowing time for sub-Doppler cooling, state preparation and detection, in a purely free-fall configuration, we anticipate a maximum pulse separation time of roughly $T_{max}= \SI{50}{\milli\second}$.

Final atom number and MOT loading time will be highly dependent on eventual experimental parameters. For a background pressure of $\SI{1e-8}{\torr}$, similar atom interferometer designs loading from a natural abundance Rb dispenser are able to achieve MOT captures of $N\approx 10^{7}$ atoms within approximately $\SI{1}{\second}$ of loading~\cite{RBMOTloading}. Using these values as a benchmark, we can estimate a shot noise limitation of $\sigma_\phi^\textnormal{QPN}=\SI{0.25}{\milli\radian}$ and a cycle time of $T_c \approx \SI{1.5}{\second}$. For a pulse separation time of $T = \SI{50}{\milli\second}$, we can estimate an acceleration noise floor of the atom interferometer of $\sigma_a^\textnormal{QPN} \simeq \SI{6.5}{\nano\meter\per\second\squared\per\sqrt{\hertz}}$.

\subsection{Optomechanical Retro-reflector Design}
The optimum OMRR design will yield the lowest hybrid sensor acceleration sensitivity for a given atom interferometer transfer function.  We can calculate this hybrid sensor acceleration sensitivity using Equation~\ref{eq:sigmasq}, where the acceleration noise power spectral density, $S_a(2 \pi n f_c)$, is taken to be the OMRR acceleration self-noise up to the sensor bandwidth. The rationale for this is that inertial noise correction using the OMRR signal can be conducted down to the OMRR self-noise over the bandwidth of the sensor. For frequencies above the OMRR bandwidth, we use the widely-used Peterson High Noise Model of general ambient seismic noise~\cite{peterson1993observations,newpeterson} as the acceleration power spectral density.

The self-noise of the OMRR is dependent on key parameters in the design of the OMRR such as, the resonance frequency, oscillator mass, and geometrical dimensions, but to first order is given by the thermal noise floor shown in Equation~\ref{eq:thermalnoisefloor}. Fixing all mechanical properties except the resonance, we optimize the hybrid sensor sensitivity as a function of OMRR bandwidth. Furthermore, we can use Equation~\ref{eq:x2a} to determine the displacement sensitivity required to resolve the self-noise at that scale. 

\begin{figure}[htbp]
\centering
\fbox{\includegraphics[width=.90\linewidth]{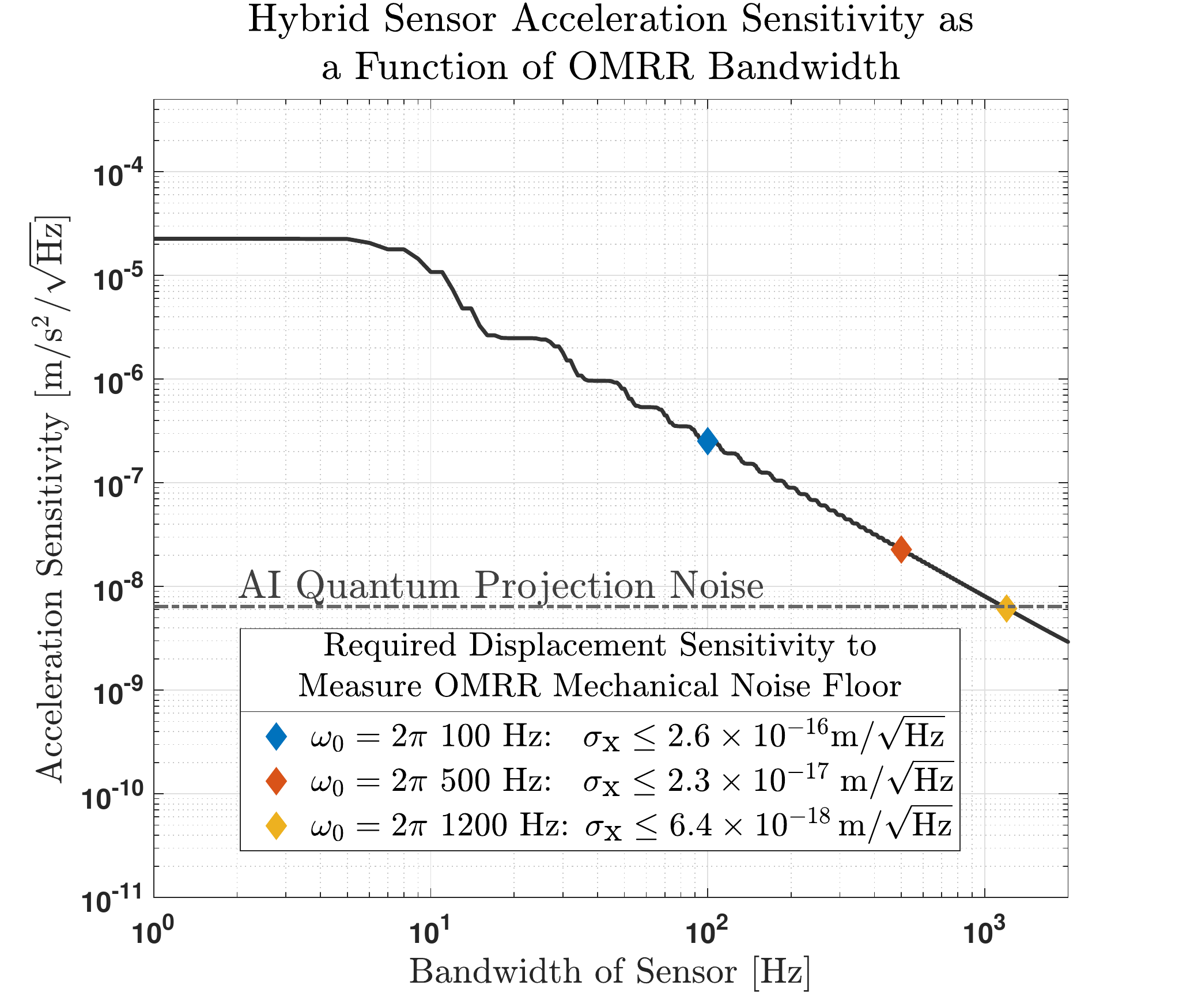}}
\caption{Calculation results of the hybrid sensor acceleration sensitivity as a function of the OMRR bandwidth. The sensitivity of the hybrid sensor at each bandwidth is determined from Equation~\ref{eq:sigmasq} with $S_a(\omega)$ being the mechanical noise floor (Equation~\ref{eq:thermalnoisefloor}) of the OMRR up to its nominal bandwidth, $\omega_0$. {The step-like features as a function of bandwidth are a result of zeros in the atom interferometer transfer function $H_\phi(\omega)$~\cite{Merlet_2009}}. For frequencies higher than resonance, the OMRR loses sensitivity and we assume the Peterson noise in this regime.  These plots were calculated for a $T=\SI{50}{\milli\second}$ atom interferometer with a cycle time of $T_c=\SI{1.5}{\second}$, an OMRR test mass $m=\SI{2}{\gram}$, and $Q = \num{5e5}$. The minimum displacement sensitivity required to measure the thermal noise floor is indicated for three OMRR resonance frequencies.}
\label{fig:bandwidthsensitivity}
\end{figure}

For this calculation, we utilize the following OMRR parameters: a) an inertial sensing test mass $m=\SI{2}{\gram}$, and b) a mechanical quality factor of $Q= \num{5E5}$; which are typical parameters from previously constructed high and low resonance optomechanical sensors~\cite{guzman2014high,OptomehanicalInertial}.  
Figure~\ref{fig:bandwidthsensitivity} shows our analysis results for the hybrid sensor sensitivity as a function of the OMRR bandwidth for a pulse separation time $T=\SI{50}{\milli\second}$ and a cycle time of $T_c = \SI{1.5}{\second}$.

\begin{figure}[htbp]
\centering
\fbox{\includegraphics[width=.9\linewidth]{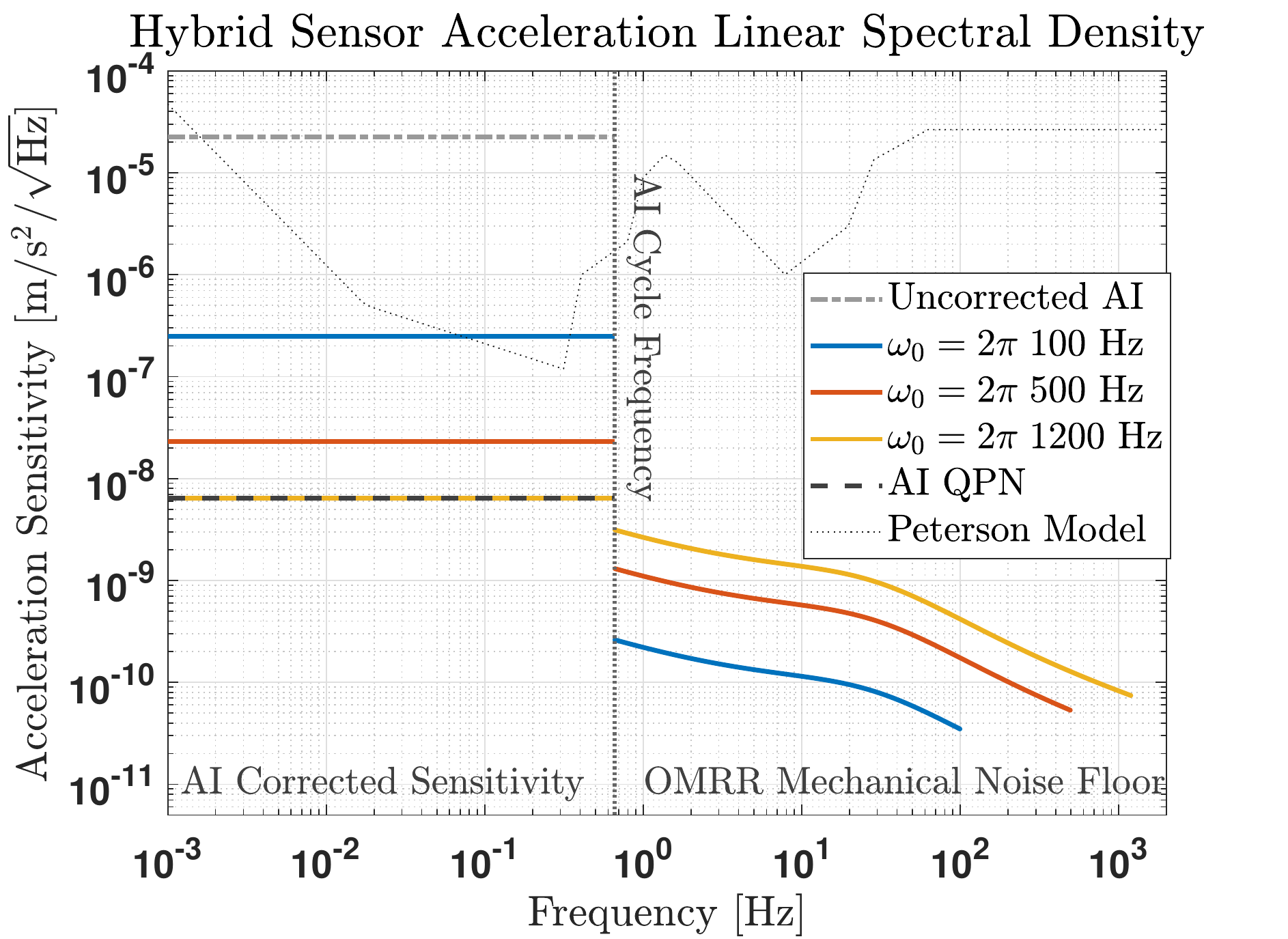}}
\caption{Linear spectral densities of the hybrid sensor acceleration sensitivity. Below the cycle frequency $f_c = \SI{.67}{\hertz}$, the OMRR signal is referenced to the atom interferometer to correct long term drifts. The improvement to the atom interferometer (below the cycle frequency) and mechanical noise floor (above the cycle frequency) is plotted for three different OMRR resonances: $\omega_0 = 2 \pi\cdot \SI{100}{\hertz}$ , $\omega_0 = 2 \pi\cdot \SI{500}{\hertz}$, $\omega_0 = 2 \pi\cdot \SI{1200}{\hertz}$. The Peterson noise floor and atom interferometer sensitivity calculated using it are also plotted as a reference.}
\label{fig:bandwidthsensitivity2}
\end{figure}

The overall acceleration linear spectral density of the hybrid sensor is given by the atom interferometer sensitivity for frequencies below the cycle rate, and above this point by the OMRR thermal noise floor. In Figure~\ref{fig:bandwidthsensitivity2}, we include the resulting quantum hybrid inertial sensor sensitivity for three OMRR example designs with different bandwidths: a) $2 \pi \cdot \SI{100}{\hertz}$ (blue), b) $2 \pi \cdot \SI{500}{\hertz}$ (red), and c) $2 \pi \cdot \SI{1200}{\hertz}$ (orange). Increasing the bandwidth of the OMRR increases the overall hybrid sensor sensitivity, however, it also significantly increases the required displacement sensitivity on the OMRR test mass dynamics, as described by Equation~\ref{eq:x2a}.  

As shown in Figure~\ref{fig:bandwidthsensitivity2}, we find that a OMRR bandwidth of  $2 \pi \cdot \SI{ 1200}{\hertz}$ is required to reach the quantum projection noise for our atom interferometer design. In order to achieve inertial measurement sensitivities with such an OMRR at its thermal noise floor, we require a displacement sensitivity of $\sigma_x \leq \SI{6.4e-18}{\meter\per\sqrt{\hertz}}$. This level of displacement sensitivity may be achievable with advanced and sophisticated laser interferometry techniques; however, implementing such a test mass displacement sensing system in a field-capable sensor is highly challenging. Higher displacement sensitivities than $\sigma_x = \SI{1e-16}{\meter\per\sqrt{\Hz}}$ over frequencies below 1\,kHz have not been reported in highly compact portable systems, but rather only demonstrated in ground-based gravitational wave detectors such as LIGO~\cite{Ligo}.

\begin{figure}[htbp]
\centering
\fbox{\includegraphics[width=.90\linewidth]{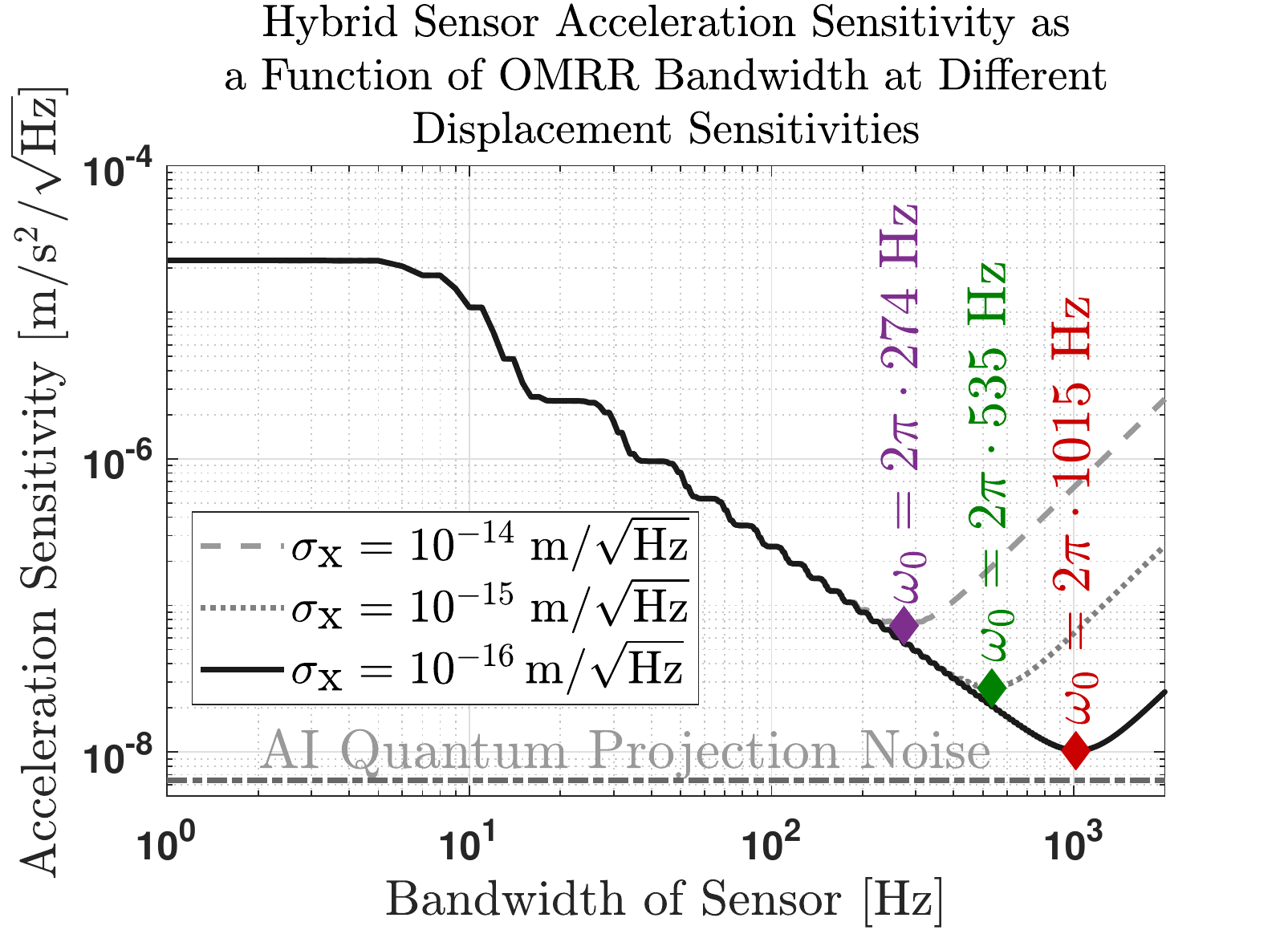}}
\caption{Calculation of the hybrid acceleration sensitivity as a function sensor bandwidth for given (achievable) displacement sensitivities: $\sigma_x = \SI{1e-14}{\meter}$, $\sigma_x = \SI{1e-15}{\meter}$ and $\sigma_x = \SI{1e-16}{\meter}$. The optimal sensor bandwidth yielding the best hybrid sensor acceleration sensitivity is indicated on each plot (solid diamond), and corresponds to a value of $2\pi\cdot\,$\SI{274}{\hertz}, $2\pi\cdot\,$\SI{535}{\hertz}, and $2\pi\cdot\,$\SI{1015}{\hertz}, respectively.  A sensor with a bandwidth $2\pi\cdot\,$\SI{1015}{\hertz} would yield an optimal sensor in line with previous optomechanical sensors which achieved displacement sensitivities of $\sigma_x \leq \SI{1e-16}{\meter\per\sqrt{\Hz}}$. Achieving higher displacement sensitivities will allow us to reach the atom interferometer quantum projection noise.}
\label{fig:dispbandy}
\end{figure}

Conversely, we can estimate the optimal hybrid sensor acceleration sensitivity for a displacement sensitivity level that can likely be realized in a portable system. Figure~\ref{fig:dispbandy} depicts the expected hybrid sensor acceleration sensitivity as a function of OMRR bandwidths and their optimums at three different OMRR test mass displacement sensitivity levels that we consider feasible to achieve: a) 274\,Hz that corresponds to $\sigma_x = \SI{1e-14}{\meter\per\sqrt{\Hz}}$ (purple), b) 535\,Hz that corresponds to $\sigma_x = \SI{1e-15}{\meter\per\sqrt{\Hz}}$ (green), and c) 1015\,Hz that corresponds to $\sigma_x = \SI{1e-16}{\meter\per\sqrt{\Hz}}$ (red).  

\begin{figure}[htbp]
\centering
\fbox{\includegraphics[width=.95\linewidth]{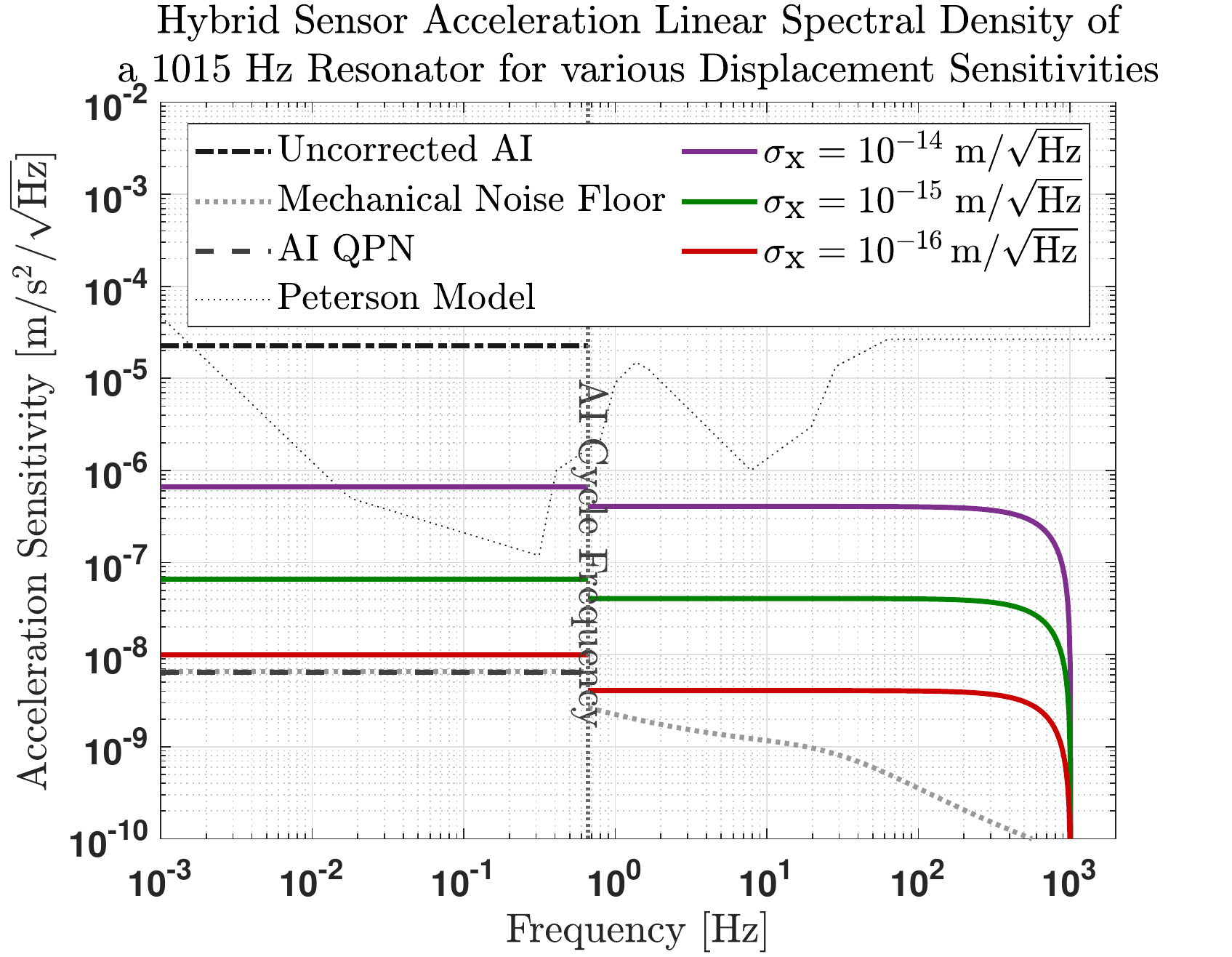}}
\caption{Acceleration linear spectral densities of the hybrid sensor for a $\omega_0 = 2\pi\cdot \SI{1015}{\hertz}$ sensor at displacement sensitivities $\sigma_x = \SI{1e-14}{\meter}$ in purple, $\sigma_x = \SI{1e-15}{\meter}$ in green, and $\sigma_x = \SI{1e-16}{\meter}$ in red, respectively. Below the cycle frequency ($f_c = \SI{0.67}{\hertz}$) the OMRR is referenced to the atom interferometer, and the atom interferometer sensitivity is calculated from Equation~\ref{eq:sigmasq}. As reference levels, we include an estimate of the uncorrected atom interferometer sensitivity (black dashed-dotted line), as well as the quantum projection noise (grey dashed line). These are calculated using the Peterson High Noise Model (grey dotted line). Above the atom interferometer cycle frequency, the displacement limitation is plotted.}
\label{fig:dispbandywhy}
\end{figure}

Previous optomechanical sensors~\cite{guzman2014high} have demonstrated displacement sensitivities of $\sigma_x = \SI{2e-16}{\meter\per\sqrt{\Hz}}$ utilizing Fabry-Perot interferometry. Taking these results as a reference, $\sigma_x \leq \SI{1e-16}{\meter\per\sqrt{\Hz}}$, we obtain an optimum trade-off between hybrid sensor sensitivity and required displacement sensitivity for a resonance of $\omega_0 = 2 \pi \cdot \SI{1015}{\hertz}$. Such a system yields a quantum hybrid inertial sensing sensitivity of $\sigma_a^\textnormal{HS} = \SI{1.02e-8}{\meter\per\second\squared\per\sqrt{\hertz}}$. The resulting acceleration sensitivity for each of the three hybrid sensor configurations outlined in Figure~\ref{fig:dispbandy}, is shown as the corresponding linear spectral densities in Figure~\ref{fig:dispbandywhy}.

\label{sec:omrrdesign}
We designed the OMRR as a drumhead mechanical oscillator that will allow us to place and quasi-monolithically attach a fixed mirror to the sensor in order to complete the Fabry-Perot cavity to be used to measure the OMRR test mass displacement. We will coat the OMRR test mass on one (the inner) side with a highly reflective layer for \SI{1560}{\nano\meter} light, as part of the Fabry-Perot optics, and on the other (outer) side to reflect the Raman beam at \SI{780}{\nano\meter} that is used for atom interferometry.

Physically, we are constrained to the dimensions of the $\SI{2.75}{\inch}$ chamber cube shown in Figure~\ref{fig:AIDiagram}, which has a bore distance of $\varnothing = \SI{38.1}{\milli\meter}$. We run numerical analysis using COMSOL to determine a sensor design of suitable dimensions for this vacuum chamber. The optimized design yielded  a cylindrical sensor with resonance $2 \pi \cdot \SI{1013.9}{\hertz}$, a frame diameter of \SI{35.5}{\milli\meter} and a height of \SI{27.9}{\milli\meter}. The test mass of the OMRR has a diameter of \SI{17.3}{\milli\meter}, thickness of \SI{3.6}{\milli\meter}, and mass of \SI{2.3}{\gram}. The test mass is supported by four flexures; each flexure has a length of \SI{7.6}{\milli\meter}, width of \SI{5}{\milli\meter}, and thickness of \SI{310}{\micro\meter}. 

\begin{figure}[htbp]
\centering
\fbox{\includegraphics[width=.95\linewidth]{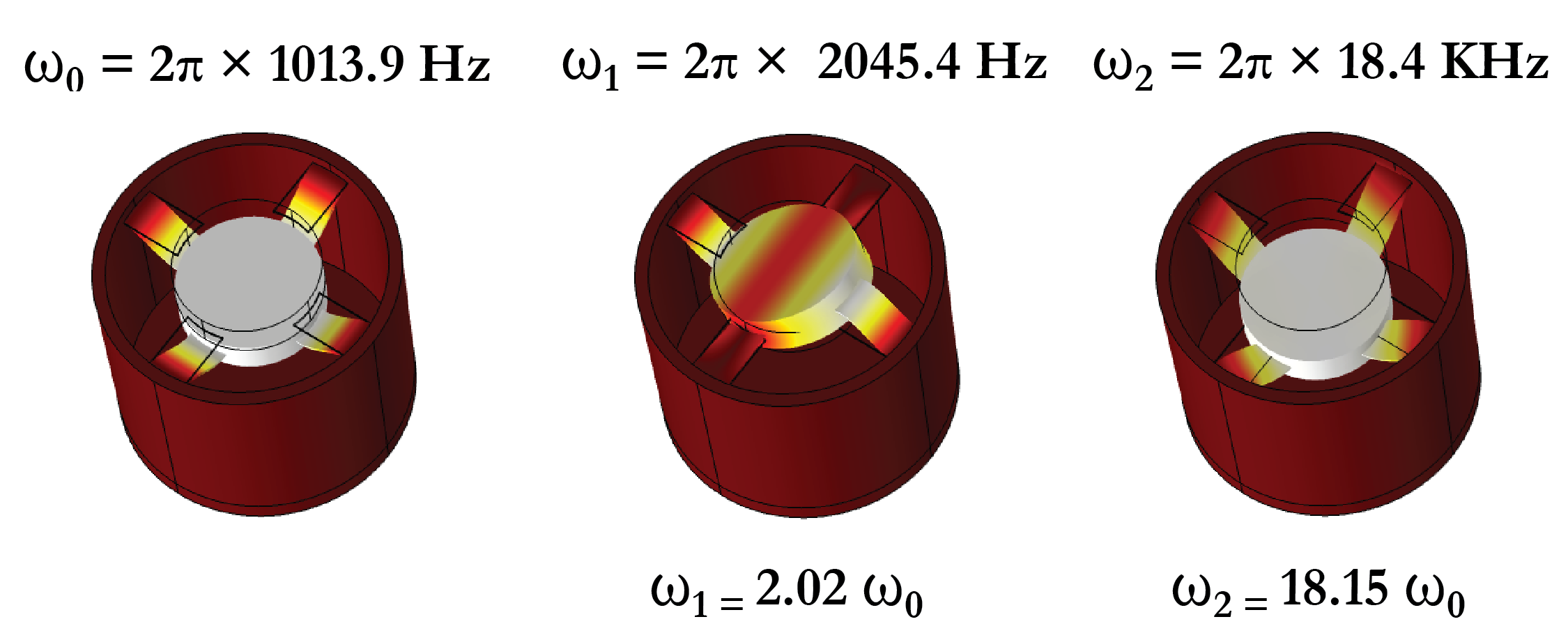}}
\caption{Modal finite element analysis of the fundamental and higher order harmonics of the OMRR. Color indicates displacement from equilibrium, with red representing no displacement, and white representing the maximum. The first harmonic is $\omega_1= 2 \pi \cdot \SI{2045.4}{\hertz}$ and the second harmonic is $\omega_2= 2 \pi \cdot \SI{18.4}{\kilo\hertz}$. Intentionally, the OMRR has been designed such that higher harmonics are high enough in frequency to avoid intermodal beat notes at frequencies at or below the fundamental resonance, $\omega_0$. }
\label{fig:Higherordermodes}
\end{figure}

We modeled the dynamics of the OMRR sensor to include bulk, surface, and thermoelastic losses~\cite{OptomehanicalInertial}, and conducted a finite element analysis to determine the first three mechanical modes, which are a) the fundamental test mass displacement mode at $\omega_0 = 2 \pi \cdot \SI{1013.9}{\hertz}$, b) a tip-tilt mode at $\omega_1 = 2 \pi \cdot \SI{2045.4}{\hertz}$, and c) a translational displacement mode at $\omega_2 = 2 \pi \cdot \SI{18.4}{\kilo\hertz}$. Figure~\ref{fig:Higherordermodes} illustrates the fundamental oscillation of the sensor, as well as these higher order modes. Under a change of acceleration of 1 $g$,  the maximum displacements for each mode are \SI{2e-7}{\meter}, \SI{6e-8}{\meter} and \SI{7e-10}{\meter} respectively. We designed the OMRR mechanics such that the frequency separation between the fundamental mode and higher order modes, as well as any intermodal beats, occur at frequencies higher than $\omega_0$ to minimize cross-talk. 

\section{Conclusion}

In this article we presented our rationale, analysis, and design considerations for a quantum hybrid inertial sensor using atom interferometers with optomechanical inertial sensing retro-reflecting mirrors. Our analysis results show that by carefully designing and implementing a well-matched OMRR to a given atom interferometer, it is possible to enhance its functionality for field deployment, yielding an inertial sensor of overall higher performance in acceleration sensitivity and observation bandwidth. Our analysis is based on previous results demonstrated by both atom interferometers and optomechanical inertial sensors and outlines our development path for a compact quantum hybrid inertial sensor that we will use in upcoming experiment studies.

From these results we obtained an OMRR sensor design with a resonance of $\omega_0= 2 \pi \cdot \SI{1013.9}{\hertz}$, which when combined with our atom interferometer design, yields an expected combined noise floor on the order of $\sigma_a^\textnormal{HS} = \SI{1e-8}{\meter\per\second\squared\per\sqrt{\hertz}}$ over a combined measurement bandwidth from DC to 1\,kHz. 

Forthcoming research activities encompass fabrication and testing of OMRR candidates, as well as integration and operation with our atom interferometer, outlined in Figure~\ref{fig:AIDiagram}. We will then perform tests of the resulting quantum hybrid inertial sensor in vibrationally noisy environments, correcting the coupling of inertial noise to the matter-wave phase using the OMRR observations, which will ultimately reduce the size and weight of the overall system by sparing the need of large and heavy vibration isolation platforms.

We expect that these improvements will increase the deployment capabilities of compact high-accuracy quantum inertial sensors in the field.

\section{Funding Information}
This work was supported, in part, by the National Geospatial-Intelligence Agency (Grant Number: HMA04761912009) and the National Science Foundation (Grant Number: PHY-1912106)

\section{Acknowledgments}
We acknowledge D. Schlippert, K. Joo and S. Nobel for helpful discussions.
The authors thank Cristina Guzman for reviewing and improving parts of the manuscript.

\bibliography{Reference}

\begin{thebibliography}{32}%
\makeatletter
\providecommand \@ifxundefined [1]{%
 \@ifx{#1\undefined}
}%
\providecommand \@ifnum [1]{%
 \ifnum #1\expandafter \@firstoftwo
 \else \expandafter \@secondoftwo
 \fi
}%
\providecommand \@ifx [1]{%
 \ifx #1\expandafter \@firstoftwo
 \else \expandafter \@secondoftwo
 \fi
}%
\providecommand \natexlab [1]{#1}%
\providecommand \enquote  [1]{``#1''}%
\providecommand \bibnamefont  [1]{#1}%
\providecommand \bibfnamefont [1]{#1}%
\providecommand \citenamefont [1]{#1}%
\providecommand \href@noop [0]{\@secondoftwo}%
\providecommand \href [0]{\begingroup \@sanitize@url \@href}%
\providecommand \@href[1]{\@@startlink{#1}\@@href}%
\providecommand \@@href[1]{\endgroup#1\@@endlink}%
\providecommand \@sanitize@url [0]{\catcode `\\12\catcode `\$12\catcode
  `\&12\catcode `\#12\catcode `\^12\catcode `\_12\catcode `\%12\relax}%
\providecommand \@@startlink[1]{}%
\providecommand \@@endlink[0]{}%
\providecommand \url  [0]{\begingroup\@sanitize@url \@url }%
\providecommand \@url [1]{\endgroup\@href {#1}{\urlprefix }}%
\providecommand \urlprefix  [0]{URL }%
\providecommand \Eprint [0]{\href }%
\providecommand \doibase [0]{http://dx.doi.org/}%
\providecommand \selectlanguage [0]{\@gobble}%
\providecommand \bibinfo  [0]{\@secondoftwo}%
\providecommand \bibfield  [0]{\@secondoftwo}%
\providecommand \translation [1]{[#1]}%
\providecommand \BibitemOpen [0]{}%
\providecommand \bibitemStop [0]{}%
\providecommand \bibitemNoStop [0]{.\EOS\space}%
\providecommand \EOS [0]{\spacefactor3000\relax}%
\providecommand \BibitemShut  [1]{\csname bibitem#1\endcsname}%
\let\auto@bib@innerbib\@empty
\bibitem [{\citenamefont {Freier}\ \emph {et~al.}(2016)\citenamefont {Freier},
  \citenamefont {Hauth}, \citenamefont {Schkolnik}, \citenamefont {Leykauf},
  \citenamefont {Schilling}, \citenamefont {Wziontek}, \citenamefont
  {Scherneck}, \citenamefont {M{\"u}ller},\ and\ \citenamefont
  {Peters}}]{freier2016mobile}%
  \BibitemOpen
  \bibfield  {author} {\bibinfo {author} {\bibfnamefont {C.}~\bibnamefont
  {Freier}}, \bibinfo {author} {\bibfnamefont {M.}~\bibnamefont {Hauth}},
  \bibinfo {author} {\bibfnamefont {V.}~\bibnamefont {Schkolnik}}, \bibinfo
  {author} {\bibfnamefont {B.}~\bibnamefont {Leykauf}}, \bibinfo {author}
  {\bibfnamefont {M.}~\bibnamefont {Schilling}}, \bibinfo {author}
  {\bibfnamefont {H.}~\bibnamefont {Wziontek}}, \bibinfo {author}
  {\bibfnamefont {H.-G.}\ \bibnamefont {Scherneck}}, \bibinfo {author}
  {\bibfnamefont {J.}~\bibnamefont {M{\"u}ller}}, \ and\ \bibinfo {author}
  {\bibfnamefont {A.}~\bibnamefont {Peters}},\ }\href@noop {} {\bibfield
  {journal} {\bibinfo  {journal} {Journal of Physics: Conference Series}\
  }\textbf {\bibinfo {volume} {723}},\ \bibinfo {pages} {012050} (\bibinfo
  {year} {2016})}\BibitemShut {NoStop}%
\bibitem [{\citenamefont {Bidel}\ \emph {et~al.}(2018)\citenamefont {Bidel},
  \citenamefont {Zahzam}, \citenamefont {Blanchard}, \citenamefont {Bonnin},
  \citenamefont {Cadoret}, \citenamefont {Bresson}, \citenamefont {Rouxel},\
  and\ \citenamefont {Lequentrec-Lalancette}}]{bidel2018absolute}%
  \BibitemOpen
  \bibfield  {author} {\bibinfo {author} {\bibfnamefont {Y.}~\bibnamefont
  {Bidel}}, \bibinfo {author} {\bibfnamefont {N.}~\bibnamefont {Zahzam}},
  \bibinfo {author} {\bibfnamefont {C.}~\bibnamefont {Blanchard}}, \bibinfo
  {author} {\bibfnamefont {A.}~\bibnamefont {Bonnin}}, \bibinfo {author}
  {\bibfnamefont {M.}~\bibnamefont {Cadoret}}, \bibinfo {author} {\bibfnamefont
  {A.}~\bibnamefont {Bresson}}, \bibinfo {author} {\bibfnamefont
  {D.}~\bibnamefont {Rouxel}}, \ and\ \bibinfo {author} {\bibfnamefont
  {M.}~\bibnamefont {Lequentrec-Lalancette}},\ }\href@noop {} {\bibfield
  {journal} {\bibinfo  {journal} {Nature communications}\ }\textbf {\bibinfo
  {volume} {9}},\ \bibinfo {pages} {1} (\bibinfo {year} {2018})}\BibitemShut
  {NoStop}%
\bibitem [{\citenamefont {Zhou}\ \emph {et~al.}(2011)\citenamefont {Zhou},
  \citenamefont {Xiong}, \citenamefont {Yang}, \citenamefont {Tang},
  \citenamefont {Peng}, \citenamefont {Hao}, \citenamefont {Li}, \citenamefont
  {Liu}, \citenamefont {Wang},\ and\ \citenamefont
  {Zhan}}]{zhou2011development}%
  \BibitemOpen
  \bibfield  {author} {\bibinfo {author} {\bibfnamefont {L.}~\bibnamefont
  {Zhou}}, \bibinfo {author} {\bibfnamefont {Z.}~\bibnamefont {Xiong}},
  \bibinfo {author} {\bibfnamefont {W.}~\bibnamefont {Yang}}, \bibinfo {author}
  {\bibfnamefont {B.}~\bibnamefont {Tang}}, \bibinfo {author} {\bibfnamefont
  {W.}~\bibnamefont {Peng}}, \bibinfo {author} {\bibfnamefont {K.}~\bibnamefont
  {Hao}}, \bibinfo {author} {\bibfnamefont {R.}~\bibnamefont {Li}}, \bibinfo
  {author} {\bibfnamefont {M.}~\bibnamefont {Liu}}, \bibinfo {author}
  {\bibfnamefont {J.}~\bibnamefont {Wang}}, \ and\ \bibinfo {author}
  {\bibfnamefont {M.}~\bibnamefont {Zhan}},\ }\href@noop {} {\bibfield
  {journal} {\bibinfo  {journal} {General Relativity and Gravitation}\ }\textbf
  {\bibinfo {volume} {43}},\ \bibinfo {pages} {1931} (\bibinfo {year}
  {2011})}\BibitemShut {NoStop}%
\bibitem [{\citenamefont {Farah}\ \emph {et~al.}(2014)\citenamefont {Farah},
  \citenamefont {Guerlin}, \citenamefont {Landragin}, \citenamefont {Bouyer},
  \citenamefont {Gaffet}, \citenamefont {Dos~Santos},\ and\ \citenamefont
  {Merlet}}]{farah2014underground}%
  \BibitemOpen
  \bibfield  {author} {\bibinfo {author} {\bibfnamefont {T.}~\bibnamefont
  {Farah}}, \bibinfo {author} {\bibfnamefont {C.}~\bibnamefont {Guerlin}},
  \bibinfo {author} {\bibfnamefont {A.}~\bibnamefont {Landragin}}, \bibinfo
  {author} {\bibfnamefont {P.}~\bibnamefont {Bouyer}}, \bibinfo {author}
  {\bibfnamefont {S.}~\bibnamefont {Gaffet}}, \bibinfo {author} {\bibfnamefont
  {F.~P.}\ \bibnamefont {Dos~Santos}}, \ and\ \bibinfo {author} {\bibfnamefont
  {S.}~\bibnamefont {Merlet}},\ }\href@noop {} {\bibfield  {journal} {\bibinfo
  {journal} {Gyroscopy and Navigation}\ }\textbf {\bibinfo {volume} {5}},\
  \bibinfo {pages} {266} (\bibinfo {year} {2014})}\BibitemShut {NoStop}%
\bibitem [{\citenamefont {Wu}\ \emph {et~al.}(2019)\citenamefont {Wu},
  \citenamefont {Pagel}, \citenamefont {Malek}, \citenamefont {Nguyen},
  \citenamefont {Zi}, \citenamefont {Scheirer},\ and\ \citenamefont
  {M{\"u}ller}}]{wu2019gravity}%
  \BibitemOpen
  \bibfield  {author} {\bibinfo {author} {\bibfnamefont {X.}~\bibnamefont
  {Wu}}, \bibinfo {author} {\bibfnamefont {Z.}~\bibnamefont {Pagel}}, \bibinfo
  {author} {\bibfnamefont {B.~S.}\ \bibnamefont {Malek}}, \bibinfo {author}
  {\bibfnamefont {T.~H.}\ \bibnamefont {Nguyen}}, \bibinfo {author}
  {\bibfnamefont {F.}~\bibnamefont {Zi}}, \bibinfo {author} {\bibfnamefont
  {D.~S.}\ \bibnamefont {Scheirer}}, \ and\ \bibinfo {author} {\bibfnamefont
  {H.}~\bibnamefont {M{\"u}ller}},\ }\href@noop {} {\bibfield  {journal}
  {\bibinfo  {journal} {Science advances}\ }\textbf {\bibinfo {volume} {5}},\
  \bibinfo {pages} {eaax0800} (\bibinfo {year} {2019})}\BibitemShut {NoStop}%
\bibitem [{\citenamefont {Hinton}\ \emph {et~al.}(2017)\citenamefont {Hinton},
  \citenamefont {Perea-Ortiz}, \citenamefont {Winch}, \citenamefont {Briggs},
  \citenamefont {Freer}, \citenamefont {Moustoukas}, \citenamefont
  {Powell-Gill}, \citenamefont {Squire}, \citenamefont {Lamb}, \citenamefont
  {Rammeloo} \emph {et~al.}}]{hinton2017portable}%
  \BibitemOpen
  \bibfield  {author} {\bibinfo {author} {\bibfnamefont {A.}~\bibnamefont
  {Hinton}}, \bibinfo {author} {\bibfnamefont {M.}~\bibnamefont {Perea-Ortiz}},
  \bibinfo {author} {\bibfnamefont {J.}~\bibnamefont {Winch}}, \bibinfo
  {author} {\bibfnamefont {J.}~\bibnamefont {Briggs}}, \bibinfo {author}
  {\bibfnamefont {S.}~\bibnamefont {Freer}}, \bibinfo {author} {\bibfnamefont
  {D.}~\bibnamefont {Moustoukas}}, \bibinfo {author} {\bibfnamefont
  {S.}~\bibnamefont {Powell-Gill}}, \bibinfo {author} {\bibfnamefont
  {C.}~\bibnamefont {Squire}}, \bibinfo {author} {\bibfnamefont
  {A.}~\bibnamefont {Lamb}}, \bibinfo {author} {\bibfnamefont {C.}~\bibnamefont
  {Rammeloo}},  \emph {et~al.},\ }\href@noop {} {\bibfield  {journal} {\bibinfo
   {journal} {Philosophical Transactions of the Royal Society A: Mathematical,
  Physical and Engineering Sciences}\ }\textbf {\bibinfo {volume} {375}},\
  \bibinfo {pages} {20160238} (\bibinfo {year} {2017})}\BibitemShut {NoStop}%
\bibitem [{\citenamefont
  {Garrido~Alzar}(2019{\natexlab{a}})}]{garrido2019compact}%
  \BibitemOpen
  \bibfield  {author} {\bibinfo {author} {\bibfnamefont {C.~L.}\ \bibnamefont
  {Garrido~Alzar}},\ }\href@noop {} {\bibfield  {journal} {\bibinfo  {journal}
  {AVS Quantum Science}\ }\textbf {\bibinfo {volume} {1}},\ \bibinfo {pages}
  {014702} (\bibinfo {year} {2019}{\natexlab{a}})}\BibitemShut {NoStop}%
\bibitem [{\citenamefont {Cheiney}\ \emph {et~al.}(2018)\citenamefont
  {Cheiney}, \citenamefont {Fouch{\'e}}, \citenamefont {Templier},
  \citenamefont {Napolitano}, \citenamefont {Battelier}, \citenamefont
  {Bouyer},\ and\ \citenamefont {Barrett}}]{cheiney2018navigation}%
  \BibitemOpen
  \bibfield  {author} {\bibinfo {author} {\bibfnamefont {P.}~\bibnamefont
  {Cheiney}}, \bibinfo {author} {\bibfnamefont {L.}~\bibnamefont {Fouch{\'e}}},
  \bibinfo {author} {\bibfnamefont {S.}~\bibnamefont {Templier}}, \bibinfo
  {author} {\bibfnamefont {F.}~\bibnamefont {Napolitano}}, \bibinfo {author}
  {\bibfnamefont {B.}~\bibnamefont {Battelier}}, \bibinfo {author}
  {\bibfnamefont {P.}~\bibnamefont {Bouyer}}, \ and\ \bibinfo {author}
  {\bibfnamefont {B.}~\bibnamefont {Barrett}},\ }\href@noop {} {\bibfield
  {journal} {\bibinfo  {journal} {Physical Review Applied}\ }\textbf {\bibinfo
  {volume} {10}},\ \bibinfo {pages} {034030} (\bibinfo {year}
  {2018})}\BibitemShut {NoStop}%
\bibitem [{\citenamefont
  {Garrido~Alzar}(2019{\natexlab{b}})}]{doi:10.1116/1.5120348}%
  \BibitemOpen
  \bibfield  {author} {\bibinfo {author} {\bibfnamefont {C.~L.}\ \bibnamefont
  {Garrido~Alzar}},\ }\href@noop {} {\bibfield  {journal} {\bibinfo  {journal}
  {AVS Quantum Science}\ }\textbf {\bibinfo {volume} {1}},\ \bibinfo {pages}
  {014702} (\bibinfo {year} {2019}{\natexlab{b}})}\BibitemShut {NoStop}%
\bibitem [{\citenamefont {Bongs}\ \emph {et~al.}(2019)\citenamefont {Bongs},
  \citenamefont {Holynski}, \citenamefont {Vovrosh}, \citenamefont {Bouyer},
  \citenamefont {Condon}, \citenamefont {Rasel}, \citenamefont {Schubert},
  \citenamefont {Schleich},\ and\ \citenamefont {Roura}}]{AINATURE}%
  \BibitemOpen
  \bibfield  {author} {\bibinfo {author} {\bibfnamefont {K.}~\bibnamefont
  {Bongs}}, \bibinfo {author} {\bibfnamefont {M.}~\bibnamefont {Holynski}},
  \bibinfo {author} {\bibfnamefont {J.}~\bibnamefont {Vovrosh}}, \bibinfo
  {author} {\bibfnamefont {P.}~\bibnamefont {Bouyer}}, \bibinfo {author}
  {\bibfnamefont {G.}~\bibnamefont {Condon}}, \bibinfo {author} {\bibfnamefont
  {E.}~\bibnamefont {Rasel}}, \bibinfo {author} {\bibfnamefont
  {C.}~\bibnamefont {Schubert}}, \bibinfo {author} {\bibfnamefont {W.~P.}\
  \bibnamefont {Schleich}}, \ and\ \bibinfo {author} {\bibfnamefont
  {A.}~\bibnamefont {Roura}},\ }\href@noop {} {\bibfield  {journal} {\bibinfo
  {journal} {Nature Reviews Physics}\ ,\ \bibinfo {pages} {1}} (\bibinfo {year}
  {2019})}\BibitemShut {NoStop}%
\bibitem [{\citenamefont {Cronin}\ \emph {et~al.}(2009)\citenamefont {Cronin},
  \citenamefont {Schmiedmayer},\ and\ \citenamefont
  {Pritchard}}]{chroninreview}%
  \BibitemOpen
  \bibfield  {author} {\bibinfo {author} {\bibfnamefont {A.~D.}\ \bibnamefont
  {Cronin}}, \bibinfo {author} {\bibfnamefont {J.}~\bibnamefont
  {Schmiedmayer}}, \ and\ \bibinfo {author} {\bibfnamefont {D.~E.}\
  \bibnamefont {Pritchard}},\ }\href {\doibase 10.1103/RevModPhys.81.1051}
  {\bibfield  {journal} {\bibinfo  {journal} {Rev. Mod. Phys.}\ }\textbf
  {\bibinfo {volume} {81}},\ \bibinfo {pages} {1051} (\bibinfo {year}
  {2009})}\BibitemShut {NoStop}%
\bibitem [{\citenamefont {Cheinet}\ \emph {et~al.}(2008)\citenamefont
  {Cheinet}, \citenamefont {Canuel}, \citenamefont {Dos~Santos}, \citenamefont
  {Gauguet}, \citenamefont {Yver-Leduc},\ and\ \citenamefont
  {Landragin}}]{cheinet2008measurement}%
  \BibitemOpen
  \bibfield  {author} {\bibinfo {author} {\bibfnamefont {P.}~\bibnamefont
  {Cheinet}}, \bibinfo {author} {\bibfnamefont {B.}~\bibnamefont {Canuel}},
  \bibinfo {author} {\bibfnamefont {F.~P.}\ \bibnamefont {Dos~Santos}},
  \bibinfo {author} {\bibfnamefont {A.}~\bibnamefont {Gauguet}}, \bibinfo
  {author} {\bibfnamefont {F.}~\bibnamefont {Yver-Leduc}}, \ and\ \bibinfo
  {author} {\bibfnamefont {A.}~\bibnamefont {Landragin}},\ }\href@noop {}
  {\bibfield  {journal} {\bibinfo  {journal} {IEEE Transactions on
  instrumentation and measurement}\ }\textbf {\bibinfo {volume} {57}},\
  \bibinfo {pages} {1141} (\bibinfo {year} {2008})}\BibitemShut {NoStop}%
\bibitem [{\citenamefont {Hauth}\ \emph {et~al.}(2013)\citenamefont {Hauth},
  \citenamefont {Freier}, \citenamefont {Schkolnik}, \citenamefont {Senger},
  \citenamefont {Schmidt},\ and\ \citenamefont {Peters}}]{gain}%
  \BibitemOpen
  \bibfield  {author} {\bibinfo {author} {\bibfnamefont {M.}~\bibnamefont
  {Hauth}}, \bibinfo {author} {\bibfnamefont {C.}~\bibnamefont {Freier}},
  \bibinfo {author} {\bibfnamefont {V.}~\bibnamefont {Schkolnik}}, \bibinfo
  {author} {\bibfnamefont {A.}~\bibnamefont {Senger}}, \bibinfo {author}
  {\bibfnamefont {M.}~\bibnamefont {Schmidt}}, \ and\ \bibinfo {author}
  {\bibfnamefont {A.}~\bibnamefont {Peters}},\ }\href@noop {} {\bibfield
  {journal} {\bibinfo  {journal} {Applied Physics B}\ }\textbf {\bibinfo
  {volume} {113}},\ \bibinfo {pages} {49} (\bibinfo {year} {2013})}\BibitemShut
  {NoStop}%
\bibitem [{\citenamefont {Zhou}\ \emph {et~al.}(2012)\citenamefont {Zhou},
  \citenamefont {Hu}, \citenamefont {Duan}, \citenamefont {Sun}, \citenamefont
  {Chen}, \citenamefont {Zhang},\ and\ \citenamefont
  {Luo}}]{PhysRevA.86.043630}%
  \BibitemOpen
  \bibfield  {author} {\bibinfo {author} {\bibfnamefont {M.-K.}\ \bibnamefont
  {Zhou}}, \bibinfo {author} {\bibfnamefont {Z.-K.}\ \bibnamefont {Hu}},
  \bibinfo {author} {\bibfnamefont {X.-C.}\ \bibnamefont {Duan}}, \bibinfo
  {author} {\bibfnamefont {B.-L.}\ \bibnamefont {Sun}}, \bibinfo {author}
  {\bibfnamefont {L.-L.}\ \bibnamefont {Chen}}, \bibinfo {author}
  {\bibfnamefont {Q.-Z.}\ \bibnamefont {Zhang}}, \ and\ \bibinfo {author}
  {\bibfnamefont {J.}~\bibnamefont {Luo}},\ }\href@noop {} {\bibfield
  {journal} {\bibinfo  {journal} {Phys. Rev. A}\ }\textbf {\bibinfo {volume}
  {86}},\ \bibinfo {pages} {043630} (\bibinfo {year} {2012})}\BibitemShut
  {NoStop}%
\bibitem [{\citenamefont {Merlet}\ \emph {et~al.}(2009)\citenamefont {Merlet},
  \citenamefont {{Gouët}}, \citenamefont {Bodart}, \citenamefont {Clairon},
  \citenamefont {Landragin}, \citenamefont {Santos},\ and\ \citenamefont
  {Rouchon}}]{Merlet_2009}%
  \BibitemOpen
  \bibfield  {author} {\bibinfo {author} {\bibfnamefont {S.}~\bibnamefont
  {Merlet}}, \bibinfo {author} {\bibfnamefont {J.~L.}\ \bibnamefont
  {{Gouët}}}, \bibinfo {author} {\bibfnamefont {Q.}~\bibnamefont {Bodart}},
  \bibinfo {author} {\bibfnamefont {A.}~\bibnamefont {Clairon}}, \bibinfo
  {author} {\bibfnamefont {A.}~\bibnamefont {Landragin}}, \bibinfo {author}
  {\bibfnamefont {F.~P.~D.}\ \bibnamefont {Santos}}, \ and\ \bibinfo {author}
  {\bibfnamefont {P.}~\bibnamefont {Rouchon}},\ }\href@noop {} {\bibfield
  {journal} {\bibinfo  {journal} {Metrologia}\ }\textbf {\bibinfo {volume}
  {46}},\ \bibinfo {pages} {87} (\bibinfo {year} {2009})}\BibitemShut {NoStop}%
\bibitem [{\citenamefont {Geiger}\ \emph {et~al.}(2011)\citenamefont {Geiger},
  \citenamefont {M{\'e}noret}, \citenamefont {Stern}, \citenamefont {Zahzam},
  \citenamefont {Cheinet}, \citenamefont {Battelier}, \citenamefont {Villing},
  \citenamefont {Moron}, \citenamefont {Lours}, \citenamefont {Bidel} \emph
  {et~al.}}]{airbornegravimetry}%
  \BibitemOpen
  \bibfield  {author} {\bibinfo {author} {\bibfnamefont {R.}~\bibnamefont
  {Geiger}}, \bibinfo {author} {\bibfnamefont {V.}~\bibnamefont {M{\'e}noret}},
  \bibinfo {author} {\bibfnamefont {G.}~\bibnamefont {Stern}}, \bibinfo
  {author} {\bibfnamefont {N.}~\bibnamefont {Zahzam}}, \bibinfo {author}
  {\bibfnamefont {P.}~\bibnamefont {Cheinet}}, \bibinfo {author} {\bibfnamefont
  {B.}~\bibnamefont {Battelier}}, \bibinfo {author} {\bibfnamefont
  {A.}~\bibnamefont {Villing}}, \bibinfo {author} {\bibfnamefont
  {F.}~\bibnamefont {Moron}}, \bibinfo {author} {\bibfnamefont
  {M.}~\bibnamefont {Lours}}, \bibinfo {author} {\bibfnamefont
  {Y.}~\bibnamefont {Bidel}},  \emph {et~al.},\ }\href@noop {} {\bibfield
  {journal} {\bibinfo  {journal} {Nature communications}\ }\textbf {\bibinfo
  {volume} {2}},\ \bibinfo {pages} {1} (\bibinfo {year} {2011})}\BibitemShut
  {NoStop}%
\bibitem [{\citenamefont {Bidel}\ \emph {et~al.}(2020)\citenamefont {Bidel},
  \citenamefont {Zahzam}, \citenamefont {Bresson}, \citenamefont {Blanchard},
  \citenamefont {Cadoret}, \citenamefont {Olesen},\ and\ \citenamefont
  {Forsberg}}]{bidel2020absolute}%
  \BibitemOpen
  \bibfield  {author} {\bibinfo {author} {\bibfnamefont {Y.}~\bibnamefont
  {Bidel}}, \bibinfo {author} {\bibfnamefont {N.}~\bibnamefont {Zahzam}},
  \bibinfo {author} {\bibfnamefont {A.}~\bibnamefont {Bresson}}, \bibinfo
  {author} {\bibfnamefont {C.}~\bibnamefont {Blanchard}}, \bibinfo {author}
  {\bibfnamefont {M.}~\bibnamefont {Cadoret}}, \bibinfo {author} {\bibfnamefont
  {A.~V.}\ \bibnamefont {Olesen}}, \ and\ \bibinfo {author} {\bibfnamefont
  {R.}~\bibnamefont {Forsberg}},\ }\href@noop {} {\bibfield  {journal}
  {\bibinfo  {journal} {Journal of Geodesy}\ }\textbf {\bibinfo {volume}
  {94}},\ \bibinfo {pages} {20} (\bibinfo {year} {2020})}\BibitemShut {NoStop}%
\bibitem [{\citenamefont {Touboul}\ \emph {et~al.}(1999)\citenamefont
  {Touboul}, \citenamefont {Foulon},\ and\ \citenamefont
  {Willemenot}}]{touboul1999electrostatic}%
  \BibitemOpen
  \bibfield  {author} {\bibinfo {author} {\bibfnamefont {P.}~\bibnamefont
  {Touboul}}, \bibinfo {author} {\bibfnamefont {B.}~\bibnamefont {Foulon}}, \
  and\ \bibinfo {author} {\bibfnamefont {E.}~\bibnamefont {Willemenot}},\
  }\href@noop {} {\bibfield  {journal} {\bibinfo  {journal} {Acta
  Astronautica}\ }\textbf {\bibinfo {volume} {45}},\ \bibinfo {pages} {605}
  (\bibinfo {year} {1999})}\BibitemShut {NoStop}%
\bibitem [{\citenamefont {Christophe}\ \emph {et~al.}(2018)\citenamefont
  {Christophe}, \citenamefont {Foulon}, \citenamefont {Liorzou}, \citenamefont
  {Lebat}, \citenamefont {Boulanger}, \citenamefont {Huynh}, \citenamefont
  {Zahzam}, \citenamefont {Bidel},\ and\ \citenamefont
  {Bresson}}]{christophe2018status}%
  \BibitemOpen
  \bibfield  {author} {\bibinfo {author} {\bibfnamefont {B.}~\bibnamefont
  {Christophe}}, \bibinfo {author} {\bibfnamefont {B.}~\bibnamefont {Foulon}},
  \bibinfo {author} {\bibfnamefont {F.}~\bibnamefont {Liorzou}}, \bibinfo
  {author} {\bibfnamefont {V.}~\bibnamefont {Lebat}}, \bibinfo {author}
  {\bibfnamefont {D.}~\bibnamefont {Boulanger}}, \bibinfo {author}
  {\bibfnamefont {P.-A.}\ \bibnamefont {Huynh}}, \bibinfo {author}
  {\bibfnamefont {N.}~\bibnamefont {Zahzam}}, \bibinfo {author} {\bibfnamefont
  {Y.}~\bibnamefont {Bidel}}, \ and\ \bibinfo {author} {\bibfnamefont
  {A.}~\bibnamefont {Bresson}}\ }(\bibinfo {year} {2018})\ pp.\ \bibinfo
  {pages} {85--89}\BibitemShut {NoStop}%
\bibitem [{\citenamefont {Richardson}\ \emph {et~al.}(2019)\citenamefont
  {Richardson}, \citenamefont {Nath}, \citenamefont {Rajagopalan},
  \citenamefont {Albers}, \citenamefont {Meiners}, \citenamefont {Schubert},
  \citenamefont {Tell}, \citenamefont {Wodey}, \citenamefont {Abend},
  \citenamefont {Gersemann} \emph {et~al.}}]{richardson2019opto}%
  \BibitemOpen
  \bibfield  {author} {\bibinfo {author} {\bibfnamefont {L.}~\bibnamefont
  {Richardson}}, \bibinfo {author} {\bibfnamefont {D.}~\bibnamefont {Nath}},
  \bibinfo {author} {\bibfnamefont {A.}~\bibnamefont {Rajagopalan}}, \bibinfo
  {author} {\bibfnamefont {H.}~\bibnamefont {Albers}}, \bibinfo {author}
  {\bibfnamefont {C.}~\bibnamefont {Meiners}}, \bibinfo {author} {\bibfnamefont
  {C.}~\bibnamefont {Schubert}}, \bibinfo {author} {\bibfnamefont
  {D.}~\bibnamefont {Tell}}, \bibinfo {author} {\bibfnamefont {E.}~\bibnamefont
  {Wodey}}, \bibinfo {author} {\bibfnamefont {S.}~\bibnamefont {Abend}},
  \bibinfo {author} {\bibfnamefont {M.}~\bibnamefont {Gersemann}},  \emph
  {et~al.},\ }\href@noop {} {\bibfield  {journal} {\bibinfo  {journal} {arXiv
  preprint arXiv:1902.02867}\ } (\bibinfo {year} {2019})}\BibitemShut {NoStop}%
\bibitem [{\citenamefont {Guzm{\'a}n~Cervantes}\ \emph
  {et~al.}(2014)\citenamefont {Guzm{\'a}n~Cervantes}, \citenamefont
  {Kumanchik}, \citenamefont {Pratt},\ and\ \citenamefont
  {Taylor}}]{guzman2014high}%
  \BibitemOpen
  \bibfield  {author} {\bibinfo {author} {\bibfnamefont {F.}~\bibnamefont
  {Guzm{\'a}n~Cervantes}}, \bibinfo {author} {\bibfnamefont {L.}~\bibnamefont
  {Kumanchik}}, \bibinfo {author} {\bibfnamefont {J.}~\bibnamefont {Pratt}}, \
  and\ \bibinfo {author} {\bibfnamefont {J.~M.}\ \bibnamefont {Taylor}},\
  }\href@noop {} {\bibfield  {journal} {\bibinfo  {journal} {Applied Physics
  Letters}\ }\textbf {\bibinfo {volume} {104}},\ \bibinfo {pages} {221111}
  (\bibinfo {year} {2014})}\BibitemShut {NoStop}%
\bibitem [{\citenamefont {Kasevich}\ and\ \citenamefont
  {Chu}(1992)}]{kasevich1992measurement}%
  \BibitemOpen
  \bibfield  {author} {\bibinfo {author} {\bibfnamefont {M.}~\bibnamefont
  {Kasevich}}\ and\ \bibinfo {author} {\bibfnamefont {S.}~\bibnamefont {Chu}},\
  }\href@noop {} {\bibfield  {journal} {\bibinfo  {journal} {Applied Physics
  B}\ }\textbf {\bibinfo {volume} {54}},\ \bibinfo {pages} {321} (\bibinfo
  {year} {1992})}\BibitemShut {NoStop}%
\bibitem [{\citenamefont {Le~Gou{\"e}t}\ \emph {et~al.}(2008)\citenamefont
  {Le~Gou{\"e}t}, \citenamefont {Mehlst{\"a}ubler}, \citenamefont {Kim},
  \citenamefont {Merlet}, \citenamefont {Clairon}, \citenamefont {Landragin},\
  and\ \citenamefont {Dos~Santos}}]{le2008limits}%
  \BibitemOpen
  \bibfield  {author} {\bibinfo {author} {\bibfnamefont {J.}~\bibnamefont
  {Le~Gou{\"e}t}}, \bibinfo {author} {\bibfnamefont {T.}~\bibnamefont
  {Mehlst{\"a}ubler}}, \bibinfo {author} {\bibfnamefont {J.}~\bibnamefont
  {Kim}}, \bibinfo {author} {\bibfnamefont {S.}~\bibnamefont {Merlet}},
  \bibinfo {author} {\bibfnamefont {A.}~\bibnamefont {Clairon}}, \bibinfo
  {author} {\bibfnamefont {A.}~\bibnamefont {Landragin}}, \ and\ \bibinfo
  {author} {\bibfnamefont {F.~P.}\ \bibnamefont {Dos~Santos}},\ }\href@noop {}
  {\bibfield  {journal} {\bibinfo  {journal} {Applied Physics B}\ }\textbf
  {\bibinfo {volume} {92}},\ \bibinfo {pages} {133} (\bibinfo {year}
  {2008})}\BibitemShut {NoStop}%
\bibitem [{\citenamefont {Itano}\ \emph {et~al.}(1993)\citenamefont {Itano},
  \citenamefont {Bergquist}, \citenamefont {Bollinger}, \citenamefont
  {Gilligan}, \citenamefont {Heinzen}, \citenamefont {Moore}, \citenamefont
  {Raizen},\ and\ \citenamefont {Wineland}}]{qpn}%
  \BibitemOpen
  \bibfield  {author} {\bibinfo {author} {\bibfnamefont {W.~M.}\ \bibnamefont
  {Itano}}, \bibinfo {author} {\bibfnamefont {J.~C.}\ \bibnamefont
  {Bergquist}}, \bibinfo {author} {\bibfnamefont {J.~J.}\ \bibnamefont
  {Bollinger}}, \bibinfo {author} {\bibfnamefont {J.}~\bibnamefont {Gilligan}},
  \bibinfo {author} {\bibfnamefont {D.~J.}\ \bibnamefont {Heinzen}}, \bibinfo
  {author} {\bibfnamefont {F.}~\bibnamefont {Moore}}, \bibinfo {author}
  {\bibfnamefont {M.}~\bibnamefont {Raizen}}, \ and\ \bibinfo {author}
  {\bibfnamefont {D.~J.}\ \bibnamefont {Wineland}},\ }\href@noop {} {\bibfield
  {journal} {\bibinfo  {journal} {Physical Review A}\ }\textbf {\bibinfo
  {volume} {47}},\ \bibinfo {pages} {3554} (\bibinfo {year}
  {1993})}\BibitemShut {NoStop}%
\bibitem [{\citenamefont {Rosi}\ \emph {et~al.}(2018)\citenamefont {Rosi},
  \citenamefont {Burchianti}, \citenamefont {Conclave}, \citenamefont {Naik},
  \citenamefont {Roati}, \citenamefont {Fort},\ and\ \citenamefont
  {Minardi}}]{rosi2018lambda}%
  \BibitemOpen
  \bibfield  {author} {\bibinfo {author} {\bibfnamefont {S.}~\bibnamefont
  {Rosi}}, \bibinfo {author} {\bibfnamefont {A.}~\bibnamefont {Burchianti}},
  \bibinfo {author} {\bibfnamefont {S.}~\bibnamefont {Conclave}}, \bibinfo
  {author} {\bibfnamefont {D.~S.}\ \bibnamefont {Naik}}, \bibinfo {author}
  {\bibfnamefont {G.}~\bibnamefont {Roati}}, \bibinfo {author} {\bibfnamefont
  {C.}~\bibnamefont {Fort}}, \ and\ \bibinfo {author} {\bibfnamefont
  {F.}~\bibnamefont {Minardi}},\ }\href@noop {} {\bibfield  {journal} {\bibinfo
   {journal} {Scientific reports}\ }\textbf {\bibinfo {volume} {8}},\ \bibinfo
  {pages} {1} (\bibinfo {year} {2018})}\BibitemShut {NoStop}%
\bibitem [{\citenamefont {L{\"u}cke}\ \emph {et~al.}(2011)\citenamefont
  {L{\"u}cke}, \citenamefont {Scherer}, \citenamefont {Kruse}, \citenamefont
  {Pezz{\'e}}, \citenamefont {Deuretzbacher}, \citenamefont {Hyllus},
  \citenamefont {Peise}, \citenamefont {Ertmer}, \citenamefont {Arlt},
  \citenamefont {Santos} \emph {et~al.}}]{beyondqpn}%
  \BibitemOpen
  \bibfield  {author} {\bibinfo {author} {\bibfnamefont {B.}~\bibnamefont
  {L{\"u}cke}}, \bibinfo {author} {\bibfnamefont {M.}~\bibnamefont {Scherer}},
  \bibinfo {author} {\bibfnamefont {J.}~\bibnamefont {Kruse}}, \bibinfo
  {author} {\bibfnamefont {L.}~\bibnamefont {Pezz{\'e}}}, \bibinfo {author}
  {\bibfnamefont {F.}~\bibnamefont {Deuretzbacher}}, \bibinfo {author}
  {\bibfnamefont {P.}~\bibnamefont {Hyllus}}, \bibinfo {author} {\bibfnamefont
  {J.}~\bibnamefont {Peise}}, \bibinfo {author} {\bibfnamefont
  {W.}~\bibnamefont {Ertmer}}, \bibinfo {author} {\bibfnamefont
  {J.}~\bibnamefont {Arlt}}, \bibinfo {author} {\bibfnamefont {L.}~\bibnamefont
  {Santos}},  \emph {et~al.},\ }\href@noop {} {\bibfield  {journal} {\bibinfo
  {journal} {Science}\ }\textbf {\bibinfo {volume} {334}},\ \bibinfo {pages}
  {773} (\bibinfo {year} {2011})}\BibitemShut {NoStop}%
\bibitem [{\citenamefont {Saulson}(1990)}]{Saulson}%
  \BibitemOpen
  \bibfield  {author} {\bibinfo {author} {\bibfnamefont {P.~R.}\ \bibnamefont
  {Saulson}},\ }\href@noop {} {\bibfield  {journal} {\bibinfo  {journal} {Phys.
  Rev. D}\ }\textbf {\bibinfo {volume} {42}},\ \bibinfo {pages} {2437}
  (\bibinfo {year} {1990})}\BibitemShut {NoStop}%
\bibitem [{\citenamefont {Hines}\ \emph {et~al.}(2020)\citenamefont {Hines},
  \citenamefont {Wisniewski}, \citenamefont {Richardson},\ and\ \citenamefont
  {Guzman}}]{OptomehanicalInertial}%
  \BibitemOpen
  \bibfield  {author} {\bibinfo {author} {\bibfnamefont {A.}~\bibnamefont
  {Hines}}, \bibinfo {author} {\bibfnamefont {H.}~\bibnamefont {Wisniewski}},
  \bibinfo {author} {\bibfnamefont {L.}~\bibnamefont {Richardson}}, \ and\
  \bibinfo {author} {\bibfnamefont {F.}~\bibnamefont {Guzman}},\ }\href@noop {}
  {\bibfield  {journal} {\bibinfo  {journal} {arXiv:2005.03456}\ } (\bibinfo
  {year} {2020})}\BibitemShut {NoStop}%
\bibitem [{\citenamefont {Rapol}\ \emph {et~al.}(2001)\citenamefont {Rapol},
  \citenamefont {Wasan},\ and\ \citenamefont {Natarajan}}]{RBMOTloading}%
  \BibitemOpen
  \bibfield  {author} {\bibinfo {author} {\bibfnamefont {U.~D.}\ \bibnamefont
  {Rapol}}, \bibinfo {author} {\bibfnamefont {A.}~\bibnamefont {Wasan}}, \ and\
  \bibinfo {author} {\bibfnamefont {V.}~\bibnamefont {Natarajan}},\ }\href@noop
  {} {\bibfield  {journal} {\bibinfo  {journal} {Phys. Rev. A}\ }\textbf
  {\bibinfo {volume} {64}},\ \bibinfo {pages} {023402} (\bibinfo {year}
  {2001})}\BibitemShut {NoStop}%
\bibitem [{\citenamefont {Peterson}(1993)}]{peterson1993observations}%
  \BibitemOpen
  \bibfield  {author} {\bibinfo {author} {\bibfnamefont {J.~R.}\ \bibnamefont
  {Peterson}},\ }\href@noop {} {\bibfield  {journal} {\bibinfo  {journal} {US
  Geological Survey}\ } (\bibinfo {year} {1993})}\BibitemShut {NoStop}%
\bibitem [{\citenamefont {Abd~el aal}(2013)}]{newpeterson}%
  \BibitemOpen
  \bibfield  {author} {\bibinfo {author} {\bibfnamefont {A.~E.-A.}\
  \bibnamefont {Abd~el aal}},\ }\href@noop {} {\bibfield  {journal} {\bibinfo
  {journal} {Pure and Applied Geophysics}\ }\textbf {\bibinfo {volume} {170}}
  (\bibinfo {year} {2013})}\BibitemShut {NoStop}%
\bibitem [{\citenamefont {Abbott}\ \emph {et~al.}(2016)\citenamefont {Abbott},
  \citenamefont {Abbott},\ and\ \citenamefont {\emph{et. al}}}]{Ligo}%
  \BibitemOpen
  \bibfield  {author} {\bibinfo {author} {\bibfnamefont {B.~P.}\ \bibnamefont
  {Abbott}}, \bibinfo {author} {\bibfnamefont {R.}~\bibnamefont {Abbott}}, \
  and\ \bibinfo {author} {\bibnamefont {\emph{et. al}}} (\bibinfo
  {collaboration} {LIGO Scientific Collaboration and Virgo Collaboration}),\
  }\href {\doibase 10.1103/PhysRevLett.116.131103} {\bibfield  {journal}
  {\bibinfo  {journal} {Phys. Rev. Lett.}\ }\textbf {\bibinfo {volume} {116}},\
  \bibinfo {pages} {131103} (\bibinfo {year} {2016})}\BibitemShut {NoStop}%
\end{thebibliography}%

\end{document}